# Information and Communications Technologies (ICT) and Pre-Service Education Professionals: A Case Study of Motivation and Knowledge

María Isabel Ponce-Escudero [a] and José Gómez-Galán [b]

**Abstract**: The importance of knowing ICT training and motivation -so relevant in today's society- which currently offers the first year college students, mostly in degrees in Education, focuses the object of interest in this study. The following targets have been proposed: [1] knowing what basic skills regarding initial instrumental knowledge presents the prospective teacher (aptitudes) and [2] knowing their motivation for the educational use of ICT in the classroom (attitudes). For this purpose a non-experimental descriptive quantitative methodology has been used, with a sample of subjects (N=282) of the Autonomous Region of Extremadura (Spain). The results show that new degree college students possess a basic knowledge of ICT alongside a highly positive motivation towards the use of these. However, it is worrying that they only show an instrumental and technical knowledge of computing and telematic tools implied in social environments, but not pedagogical ones. Also they are unfamiliar with the true power of social, economic, political, ethical, influence as well as the effects and problems that their misuse can generate in their future students (addictions, manipulation, consumerism, etc.). Dimensions therefore for which they are urged to be trained at University for a proper performance as future professionals in education.

---

[a] Consejería de Educación y Empleo del Gobierno de Extremadura, SEXPE (Spain); [b] Universidad Metropolitana (Puerto Rico, United States), Universidad de Extremadura (Spain) Correspondence: María Isabel Ponce-Escudero, Coordinadora de Formación, Servicio Extremeño de Empleo Público, Ayuntamiento de Herrera del Duque, 06670 Badajoz (Spain). mariaisabelponce@academia.edu.

## 1. Introduction

### 1.1. Problem of research

We live in a society exposed to continuous changes and technological developments in which ICT are constantly increasing their relevance and power. Each of the sectors of our life (social, labor, economic, etc.) is exposed to these changes and developments, which in turn affect the field of education, which is considered to be one of the cornerstones of our society.

ICT play a decisive role in the process of teaching-learning and this is reflected as such in scientific literature, which details the great level of interest shown in the study of basic skills presented by these students.

It is clear in today's world that students must be technologically literate. It is a requirement which enables access to knowledge and allows them to navigate an ocean of information from which they can search, select and extract, etc., freely, autonomously and critically.

It is also essential to examine the student's motivation since it is a key factor in learning. Today it is clear that a motivated student is one who is pre disposed to learn. We agree with authors who maintain that one of the key advantages of the educational use of technology and media in the classroom is motivation, which is even more important than the ability to convey information and provide suitable learning content.

These therefore are the two main points of interest of the research performed. Focus will be placed on the initial training of future professionals in education, since it is on them that the education of new generations will depend because we will live in a world increasingly influenced by ICT; one in which both the basic training in these technologies and the right motivation for their implementation will be very important in establishing a new structure of the Teacher Training Curriculum in which ICT should have an important presence and future teachers should be trained according to the needs of a digital society.

### 1.2. Research focus

ICT in education are given preferential consideration by the educational and scientific community. Currently, researchers are in a fruitful and intense period in both theoretical production and field research, which is reflected in the remarkable growth that has occurred in recent years in the number of publications (books, articles in scientific journals, dissertations, etc.) which have focused on this field of knowledge. And one of the main lines of research in education science, which can no longer be considered new and innovative, is the verification of the initial training possessed by Teaching Studies students.

These students, as future specialists and education professionals have to be trained in the control and use of ICT, as they will become within the educational community the main agent responsible for the implementation of their educational use and contribution to their full integration in schools (Tondeur, Van Braak, Sang, Voogt, Fisser & Ottenbreit-Leftwich, 2012; Rogers & Twiddle, 2013; Fluck & Dowden, 2013). Their training is crucial for the utilization of ICT in the classroom in an effective way (Gómez-Galán & Mateos, 2002 and 2004; Selwyn, 2006; Dawson, 2008; Peeraer & Van Petegem, 2012; Chai, Koh & Tsai, 2013).

The teacher will provide students with their ICT training foundation, which will be decisive for its proper performance in accordance with new social needs. Their formation is one of the most significant challenges in education in relation to the new social, economic and cultural context represented by the information society. We have always believed that such training must start through an elementary process, seeking the establishment of a solid foundation from which access to new technologies from a pedagogical perspective is facilitated (Soler-Costa, 2011). The role of education and the digital literacy is essential in the knowledge-based society (Juszczyk, 2006).

We suggest that the training a Primary Teaching student or prospective future teacher acquires must be primarily formative and reflective, in addition to being technical and instrumental, enabling them as such to make critical use of ICT as part of their environment (Rogers & Twiddle, 2013). It is important to stress this because technology is not only education, and students may be encountered who have not had this kind of training or any other preparation of an inadequate and/or insufficient nature.

This training is critical, especially in a society that is manipulated by the major power elite that dominate communication processes, especially large media corporations. ICT -especially the Internet and media with a great presence in the Western world like television- have the power to create a parallel, virtual reality, which can distort true reality. They can distort information in a subtle, undetectable manner, thanks to the sophistication of the digital paradigm. We are under the control of what Echevarría (2007) called the *lords of air*, controlling all communication flow in the third environment.

Students at all levels of education have to be trained to select and interpret information critically, and filter the vast amount of information, all of which is not necessarily competently treated. Sometimes, and paradoxically, an excess of information can cloud the underlying meaning of the data obtained. Quantity is imposed on quality, leading -and we agree with the technologist Lanier (2006)- to a global network and Internet which is uninformed and tedious. It should not be forgotten that we live in a changing society and for this reason training during our lives is essential (Montaser, Mortada & Fawzy, 2011; Kim, Choi, Han & So, 2012).

Also it is necessary to equip schools with relevant, technological and audiovisual material to enable students to acquire a solid background. Education Colleges, at which future teachers are now being trained, are slowly paying more attention to this, as it is clear that it plays a crucial role in the acquisition of ICT training. These are college students who will train and educate citizens, who will constitute the workforce of the future, the main creators and consumers of ICT (UNESCO, 2002).

We agree with the classic Delors report (1996) that improving the quality of education must begin through the improvement of training. If we understand ICT training as a key skill that students must develop during their learning process in the school, it is necessary to ascertain whether students have acquired it properly. And in this sense it is crucial to determine what level of digital competence they have when they first arrive at university.

On the other hand, motivation is an essential factor in student learning (Elliot & Dweck, 2005; Schunk & Zimmerman, 2008; Grosskopf, 2009; Akbiyik, 2010). It is one of the engines of learning as it promotes activity and thought. Therefore it can be said that a motivated student will be more willing to learn than one who is not. Domínguez-Rodríguez & Cañamero (2008) stresses the importance of motivation and emotion in learning in its study of educational research with college students.

One of the keys to capturing the attention and interest of students, and as a consequence facilitating their learning, is the introduction of ICT in the classroom. These are considered key tools in improving learning construction (Petrauskiene & Volungeviciene, 2006; Albirini, 2006; Mueller, Wood, Willoughby, Ross & Specht, 2008; Hiralaal, 2013). They enable work to be done more dynamically through the promotion of interaction which in turn encourages an active and participatory student spirit which negates passive presentation. Martín, Beltrán & Pérez (2003) considers ICT's to be entertaining and attractive media, which draw students´ attention on their own.

It is important to note that a student's learning and motivation is not so much the tools used by the teacher but the way in which they are introduced in the classroom. Means alone cannot produce learning. It is more than teachers delegating their educational capacity as if it were a machine or tool, regardless of the fact that their work is enhanced by those aids. When used in the classroom, these tools will facilitate the teachers´ work and free up more time to build the motivation potential of students (Cox, 1999).

One of the main themes of research in recent years has focused on the power of ICT tools as student motivators in the teaching-learning process. It is a proven fact that more and more educational institutions and universities are seeking ways to introduce these technologies into the classroom because of the high level of student interest in ICT. Teaching professionals are being encouraged to include them as a priority in teaching resources. This may be further demonstrated by Gómez-Galán (2007, 2011 and 2015), López-Meneses & Gómez-Galán (2010), and Durand & Bombelli (2012).

We also find contributions that demonstrate that ICT facilitate students´ learning (Kennedy, Judd, Churchward, Gray, & Krause, 2008). This study shows that a high percentage of Australian first-year students (84 %) said that a mobile phone could assist them in their studies. These results highlight the positive attitude of students with regard to the role of ICT in the provision of educational aid.

**2. Methodology**

2.1. General background of research

The principal goal of our study is informed of the basic competence and ICT motivation featured by first-year students in Elementary and Primary School Education degrees (University of Extremadura, only this university exists in Extremadura, Spain) in order to [1] determine the initial conditions of future education professionals in relation to these technologies and means, and in parallel [2] to assess quality in college teacher education, aligning these important elements of our society, against a further study to be undertaken after graduation in which the same dimensions will be analyzed. In this research we start from a general hypothesis, in which we establish that pre-service education professionals at Extremadura (Spain), studying first year Education degrees, possess a significant initial knowledge of ICT as well as a positive motivation towards them.

2.2. Methods and participants

In this study the model of educational research has been chosen which is based on quantitative methodology, descriptive through surveys, and referred to as "non-experimental". This was considered to be ideally suited to the achievement of the aims pursued and the verification or refutation of our working hypothesis.

The study population is made up of first-year University students enrolled at the College of Education at the University of Extremadura in Elementary and Primary Education Degrees. 282 subjects were studied of which 171 belong to Primary and 111 to Elementary Education. The sample group was selected using a non-probability accidental sampling, causal or convenience, as the individual selection criteria depended upon their accessibility.

2.3. Instrument and procedures

For data collection we have used a questionnaire developed from validated questionnaires by Gómez-Galán (2003) and Pino & Soto (2010). The type of questionnaire is self-administered and individual. It consists of 24 questions divided into the following parts: user instructions, socio-demographic data and the questionnaire body (items, Likert-type scales). The reliability of the scale determined through Cronbach's alpha: $\alpha = (\kappa/(\kappa-1)) \cdot (1 - (\sum \sigma^2_i / \sigma^2_{sum}))$. Result of Cronbach's alpha= 0.871. Thus, the instrument has demonstrated strong internal consistency through this test.

In order to plan our research accurately and for it to be executed in an orderly manner four work patterns were established: [1] the first being the building of the theoretical framework of the study, for which a comprehensive scientific literature review on the subject was conducted. [2] Second, the instrument used for collecting the said data was chosen. The use of a validated questionnaire was selected, one which suited the object of study, and enabled us to proceed to the third pattern of data collection in a reasonable amount of time [3] In the application of the questionnaire three phases were developed: [3.1] *Awareness*: An e-mail was sent to university teachers of the first year of Primary and Elementary Education Degrees specifying the date, place and time when questionnaires were to be conducted. [3.2] *Questionnaire Application*: the delivery of the questionnaires was proceeded with, and their anonymity ensured. They were completed individually by the subjects being studied [3.3] *Return of results*: The results were offered by e-mail to the Faculty interested in obtaining information for the purposes of the study, the said results being helpful in the planning of their academic syllabus, especially as it was centered on and involved the use of ICT. [4] As a fourth and final pattern, analysis and interpretation of the results was proceeded with. The data was processed to Excel spreadsheet and exported to SPSS 18.0 for Windows, which enabled the statistical analysis of the results.

### 3. Results

Our sample group consisted of 282 students, of whom the majority is female (a total of 214 women compared to 68 men).They are mostly under 20 or between 20 and 30 years of age, 49.6% and 47.2 % respectively, with subjects over 30 years in the minority (3.2%), more than half of those who were surveyed are studying Primary Education (60.6%) and 39.4 % Elementary Education. The great majority belong to the social middle class (79.1%).

The evaluation on ICT-related knowledge made by future teachers is generally positive, with special skills being noted in programs of personal interaction like WhatsApp, e-mail and social networks (Facebook, Twitter, etc.), in which the majority deem themselves to be highly skilled (72.3% ) or fairly skilled (22.7%). This positive self-assessment must be added to the correct management of the e-learning platform of University of Extremadura (Moodle), where more than half of the university students are considered to be fairly skilled (55.3%), or highly skilled (34 %).

Just over half the students also consider themselves to have good management of basic programs (such as word processing and slide presentation) 50.7% and in the use of web browsers (Google, Yahoo, Bing, etc.) 50.4%. 37.9% consider themselves to be very skilled at basic programs and 38.4 % in the use of search engines on the web. 39.4% consider themselves to be fairly skilled at managing blogs and highly skilled in chats and forums (19.1%).

The results are less positive when asked about educational field tools such as author's programmes (Clic, J-Clic, Hot Potatoes, etc.) and Internet guided search activities (Webquest), as well as multimedia devices (overhead, webcams, etc.). Almost half of the students surveyed (47.5%) consider that they have some management of educational author's programmes, 15.2 % have fairly good management, and only 2.8 % of respondents rated their management of these programs as good. Just over half of respondents (50.8%) considered that they have some management of Webquest, 18.4 % fairly good management and only 5.7 % rated their management as being good. With regard to their assessment of their management of multimedia devices, the highest percentage believe they have some management (41.5%), followed by those who rate their management as being fairly good (35.8%) while the option of good management was chosen by (13.8%), the only ICT tools in which all of the students consider themselves to have good management, to a greater or lesser extent, are programmes of personal interaction and the virtual learning platform.

|  | *Percentage* | | | | Total | Average |
|---|---|---|---|---|---|---|
|  | 1 | 2 | 3 | 4 |  | (0-4) |
| Basic Programs | ,4 | 11,0 | 50,7 | 37,9 | 100,0 | 3,26 |
| Interpersonal Relations Programs | 0 | 5,0 | 22,7 | 72,3 | 100,0 | 3,67 |
| E-learning Platform | 0 | 10,6 | 55,3 | 34,0 | 100,0 | 3,23 |
| Author´s Education Programs | 34,4 | 47,5 | 15,2 | 2,8 | 100,0 | 1,87 |
| Search the Network | ,7 | 10,6 | 50,4 | 38,3 | 100,0 | 3,26 |
| Multimedia Devices | 8,9 | 41,5 | 35,8 | 13,8 | 100,0 | 2,55 |
| Guided Internet Searching | 25,2 | 50,7 | 18,4 | 5,7 | 100,0 | 2,05 |
| Blogs, Chats and Forums | 7,1 | 34,4 | 39,4 | 19,1 | 100,0 | 2,71 |

Table 1. *Knowledge. Key: 1=None; 2= Somewhat; 3= Quite; 4=Much*

As regards attitudes and motivations, overall, students agreed, more or less directly, with positive statements towards ICT, showing an interested attitude towards them. Most respondents agree that they strongly agree that ICT are essential in today's society, 59.6%, and in good agreement 34.4%. More than half of respondents strongly agree that knowing how to use computers and the Internet will be essential to continue studying and, in the future, to find a quality job, 57.4%, and in good agreement 36.9%. Almost half of the students also agree that they strongly agree that ICT are good for their academic training, 49.6%, and in good agreement 42.9 %. 41.8% of respondents are very interested in what they can do with computers and the Internet, or fairly interested, 48.6%. 37.9 % of students said they strongly agreed that ICT are an instrument to assist themselves in their learning process, or fairly interested 51.1%. The statement in which we found a low percentage of students who strongly agree (20.9%) although partially agree (55%), is the one in which it is stated that the Internet offers countless learning opportunities, with very few inconveniences, as well as being a tool designed for education. The statements in which all of the students slightly, partially or fully agree are two: ICT are good for their education and learning how to use them properly is essential to continue studying and finding a good job.

|  | *Percentage* | | | | *Total* | *Average* |
|---|---|---|---|---|---|---|
|  | 1 | 2 | 3 | 4 |  | (0-4) |
| I am interested in everything connected with computers and the Internet | 1,1 | 8,5 | 48,6 | 41,8 | 100 | 3,31 |
| ICT are good for my education | 0 | 7,4 | 42,9 | 49,6 | 100 | 3,42 |
| ICT help in my learning process | ,7 | 10,3 | 51,1 | 37,9 | 100 | 3,26 |
| Learning to use computers and the Internet is essential to study and get a job | 0 | 5,7 | 36,9 | 57,4 | 100 | 3,52 |
| ICT are essential in today's society | ,4 | 5,7 | 34,4 | 59,6 | 100 | 3,53 |
| Internet provides learning opportunities with few drawbacks. | 2,1 | 22,0 | 55,0 | 20,9 | 100 | 2,95 |

Table 2. *Reasons for using ICT. Key: 1=None; 2= Somewhat; 3= Quite; 4=Much*

Almost half of the surveyed students (48.6 %) believe that the Faculty in which they are enrolled has adequate audiovisual equipment and technology to provide proper training in ICT; however, 29.1 % disagree with that assessment. The rest of the students chose not to answer (DK/NA/REF: don't know, not available or refusal) as to whether their institution has such material, 22.3%.

|  |  | *Frequency* | *Percentage* | *Valid Percentage* | *Cumulative Percentage* |
|---|---|---|---|---|---|
| Valid | No | 82 | 29,1 | 29,1 | 29,1 |
|  | DK/NA/REF | 63 | 22,3 | 22,3 | 51,4 |
|  | Yes | 137 | 48,6 | 48,6 | 100,0 |
|  | Total | 282 | 100,0 | 100,0 |  |

Table 3. *Your faculty possesses sufficient audiovisual and technological material to acquire adequate training in ICT.*

More than half of college students (63.5%) believe they have adequate and timely training, which enables them to make critical use of ICT within their respective environments. On the other hand 25.2% consider their training to be inadequate or insufficient 5.7% of the subjects feel they are not receiving or have not had training at all. This 5.7% represents those individuals, who believe they are more fully skilled, i.e., and have already had suitable training.

|  |  | *Frequency* | *Percentage* | *Valid Percentage* | *Cumulative Percentage* |
|---|---|---|---|---|---|
| *Valid* | Not possess | 16 | 5,7 | 5,7 | 5,7 |
|  | Inadequate | 71 | 25,2 | 25,2 | 30,9 |
|  | Fair | 179 | 63,5 | 63,5 | 94,3 |
|  | Great | 16 | 5,7 | 5,7 | 100,0 |
|  | Total | 282 | 100,0 | 100,0 |  |

Table 4. *Your faculty possesses sufficient audiovisual and technological material to acquire adequate training in ICT.*

With regard to their use of these tools, the instruments most commonly used by most of the students surveyed are personal interaction programmes and 74.8% indicate that these are used extensively by them. Educational programmes 46.1% and Webquests 42.2% are those which are least used by college students. The only tools that are used to a greater or lesser extent (slightly, fairly or broadly) are personal interaction and basic programs and the e-learning platform of University of Extremadura.

|  | *Percentage* | | | | Total | Average |
|---|---|---|---|---|---|---|
|  | 1 | 2 | 3 | 4 |  | (0-4) |
| Basic Programs | 0 | 4,6 | 44,7 | 50,7 | 100 | 3,46 |
| Interpersonal Relations Programs | 0 | 7,1 | 18,1 | 74,8 | 100 | 3,68 |
| E-learning platform | 0 | 10,6 | 45,0 | 44,3 | 100 | 3,34 |
| Author´s Education Programs | 46,1 | 45,4 | 7,8 | ,7 | 100 | 1,63 |
| Search the Network | 1,8 | 6,0 | 45,4 | 46,8 | 100 | 3,37 |
| Multimedia Devices | 20,2 | 48,9 | 21,3 | 9,6 | 100 | 2,20 |
| Guided Internet Searching | 42,2 | 37,6 | 15,2 | 5 | 100 | 1,83 |
| Blogs, Chats and Forums | 16,3 | 36,9 | 31,2 | 15,6 | 100 | 2,46 |

Table 5. *Use ICT. Key: 1=None; 2= Somewhat; 3= Quite; 4=Much*

With regard to the inferential analysis, after calculating the Chi square test $\chi^2 = \Sigma \ ( \ (f_o - f_e)^2 \ / \ f_e \ )$ and contingency coefficient at the junction of nominal and scale variables, no significant differences in knowledge, use of ICT or motivation features by the students surveyed, were noted, regardless of age, gender, specialty, or social class.

71

## 4. Interpretation and Discussion

To simplify the presentation and discussion of the interpretation of the data, in seeking conclusions, the different dimensions of study will be established. First, we center on the knowledge and motivation in ICT by university students. Overall responses of college students reveal significant knowledge of ICT, most believe they correctly manage personal interaction programs (e-mail, social networks, Whatsapp, etc.) as well as the virtual teaching platform of the University of Extremadura. More than half of those surveyed consider their management of basic programs and web browsers to be sufficient. With regard to their use of these tools we find that the means most commonly used by university staff are interrelated and basic programs, especially the first mentioned. Educational programs and Webquest are tools that are used less. All of this indicates that ICT tools are used in their daily lives primarily for entertainment and leisure, and for the purposes of media and social relations (see e.g. Gómez-Galán, 2007 and 2011; Pećanac, Lambić & Marić, 2011; Pattaro, 2015).

Most undergraduate students are motivated and show great interest in ICT. They are aware of their importance in our society. They know that their operation with computers is essential in allowing them to continue studying and finding a job (cf. Albirini, 2006; Petrauskiene & Volungeviciene, 2006; Kim, Choi, Han, & So, 2012; Hortovanyi & Ferincz, 2015). They see the learning process as being facilitated by this technology but their assessment of Internet as being a tool designed to aid education without it having any drawbacks is seen as a dangerous. Thorough training in this regard towards the proper use of Internet by future education professionals is necessary as the interpretation that it has been designed for training is erroneous (Gómez-Galán, 2003 and 2011).

It is also significant that they are divided on the adequacy of audiovisual equipment present at the college in which they are enrolled as well as the appropriate technology being available to enable good training in ICT to be carried out. Such doubts suggest that technologies are under-resourced or under-utilized (cf. Barton & Haydn, 2006; Goktas, Yildirim & Yildirim, 2009; Murray & Rabiner, 2014).

Another disturbing response is that more than half of the students believed they have received adequate and timely training to enable effective use of ICT as part of their environment. This indicates in direct relation to the response given on the educational nature of the Internet, their ignorance of the nature of current communication processes, in which powerful media groups have acquired so much economic, social and political influence as well as all the accompanying pedagogical implications (analysis of advertising, consumerism, addiction potential, etc.). Training regarding this point is required (see Gómez-Galán & Lacerda, 2012; Buckingham, 2013;

Underwood, Parker & Stone, 2013; Ballesta, Lozano, Cerezo & Ayala, 2015).

On the other hand, we detect the insufficient previous training being given to university students. In ICT gaps have been found in the management of educational authors' programmes, guided Internet search activities and multimedia devices, where a large percentage of students confess to not having any management capacity and with only a very small percentage deeming themselves to be highly skilled. Educational programs and Webquest are the least used of these tools.

These results have led us to conclude that there is only one underlying reason which can explain the fact that education environmental tools are practically unknown to Teaching Degree students, bearing in mind that the initial training of newly admitted students to the College is being researched - i.e., the training received in their earlier stages of education needs to be assessed- given, it would seem, that such tools are not used by Primary and Secondary/High School teachers as part of teaching content in the classroom (see also Gallego & Alonso, 2000, Van Braak, Tondeur & Valcke, 2004; Sime & Priestley, 2005; Drent & Meelissen, 2008; Leask & Pachler, 2013; Hammond, 2014; Aesaert, Van Nijlen, Vanderlinde, Tondeur, Devlieger & van Braak, 2015).

The lack of prior training, therefore, designed specifically for ICT education, implies the need to have this introduced during their attendance at university. Focus will be required to be placed on technical and instrumental training in specific teaching programmes which as future teachers, they will need to be trained in to enable their understanding and use as teaching resources -and further enhance the use and integration of these tools to achieve specific learning goals- (cf. Haydn & Barton; 2007; Goktas, Yildirim & Yildirim, 2008; Khan & Hasan, 2013; Tondeur, Roblin, van Braak, Fisser & Voogt, 2013; Fluck & Dowden, 2013, Barak, 2014).

The information obtained in this study, which reveals gaps in technical and instrumental training of educational hardware and software and others in critical knowledge of the information society, is something that university professors and managers of education colleges will need to know in order to fill these important gaps during initial teacher training.

Finally, the results of the Chi square test must be interpreted as well as the contingency coefficient at the junction of nominal scale variables. It has been previously mentioned that no differences in sex, age, social class and specialty of respondents regarding their ICT skills and their motivation to use them were found. Results which show that today all citizens (regardless of gender, age, social position, etc.) are familiar with ICT. This comes as no surprise in a society where all areas of our lives are directly or indirectly influenced by them, as is its relevance as a means of communication, access to information, knowledge, entertainment, leisure, etc. Their power of social penetration is so great that even the most socially or financially disadvantaged citizens make frequent use of the latest ICT, either through private or public employment (or through educational institutions, libraries, knowledge centers, etc.).

## 5. Conclusions

The first and main conclusion that can be drawn is that the specific aims pursued in this research have been achieved, and confirmation of our working hypothesis confirmed. It can be said therefore, that students of Education degrees (pre-service education professionals) at the Autonomous Community of Extremadura (Spain) begin their training with significant technical and instrumental knowledge of ICT, but basically those that are inclined towards communication and entertainment/leisure systems. Likewise, they display positive motivation towards them. They have, in this sense, both the skills and appropriate attitudes towards technology and media. However, they lack a deeper understanding of the true nature of the tools used and their real social dimension, which will assume vital importance in the education of future students. This lack of understanding is also apparent in the non-recognition of the problems that are associated with unsuitable use derived addictions, manipulation, consumerism, etc., they offer no criticism or reflection, and are ignorant of the crucial dimension of their economic, political and social impact. Nor have they adequate knowledge on how to use them as teaching resources, and harbor doubts that Colleges of Education possess sufficient and proper training resources.

It is necessary, therefore, that the syllabus of initial teacher training includes pedagogical software and hardware, and that the issue of ICT is also analyzed as an object of study, from a critical standpoint, keeping the future training of those who will be its students in mind. We believe that these results are very important when one considers that both teacher training and motivation are keys to the achievement of full integration of ICT into educational institutions and comprehensive training of citizens.

## References

Aesaert, K., Van Nijlen, D., Vanderlinde, R., Tondeur, J., Devlieger, I., & van Braak, J. (2015). The Contribution of Pupil, Classroom and School Level Characteristics to Primary School Pupils' ICT Competences: A Performance-based Approach. *Computers & Education*, 87, 55-69. doi: 10.1016/j.compedu.2015.03.014

Akbiyik, C. (2010). Can Affective Computing Lead to More Effective Use of ICT in Education? *Revista de Educación*, 179-202.

Albirini, A. (2006). Teachers' Attitudes toward Information and Communication Technologies: The Case of Syrian EFL Teachers. *Computers & Education*, 47, 373-398. doi: 10.1016/j.compedu.2004.10.013

Ballesta F. J., Lozano, J., Cerezo, M. C., & Ayala, E. S. (2015). Internet, Redes Sociales y Adolescencia: Un Estudio en Centros de Educación Secundaria de la Región de Murcia. *Revista de la Facultad de Ciencias de la Educación*, 16, 109-130.

Barak, M. (2014). Closing the Gap between Attitudes and Perceptions about ICT-Enhanced Learning among Pre-Service STEM Teachers. *Journal of Science Education and Technology*, 23 (1), 1-14. doi: 10.1007/s10956-013-9446-8. doi: 10.1007/s10956-013-9446-8

Barton, R., & Haydn, T. (2006). Trainee Teachers' Views on What Helps them to Use Information and Communication Technology Effectively in Their Subject Teaching. *Journal of Computer Assisted Learning*, 22, 257-272. doi: 10.1111/j.1365-2729.2006.00175.x

Buckingham, D. (2013). *Media Education: Literacy, Learning and Contemporary Culture*. London: John Wiley & Sons.

Chai, C. S., Koh, J. H. L., & Tsai, C. C. (2013). A Review of Technological Pedagogical Content Knowledge. *Educational Technology & Society*, 16, 31-51. doi: 10.1037/t14052-000

Cox, M. J. (1999). Motivating Pupils Through the Use of ITC. In M. Leask & N. Pachler. *Learning to Teach Using ICT in the Secondary School* (pp. 19-35). London-New York: Routledge.

Dawson, V. (2008). Use of Information and Communication Technology by Early Career Science Teachers in Western Australia. *International Journal of Science Education*, 30(2), 203–219. doi: 10.1080/09500690601175551

Delors, J. (1996). *Learning: The Treasure Within. Report to UNESCO of the International Commission on Education for the Twenty-first Century*. Paris: UNESCO.

Domínguez-Rodríguez, E. & Cañamero, P. (2008). Perfil del Alumnado Extremeño de Educación Superior. *Alcántara*, 69, 49-73.

Drent, M., & Meelissen, M. (2008). Which Factors Obstruct or Stimulate Teacher Educators to use ICT Innovatively? *Computers & Education*, 51 (1), 187-199. doi: 10.1016/j.compedu.2007.05.001

Durand, P. & Bombelli C.E. (2012). El Uso de Blogs en la Formación Universitaria. *Quaderns Digital*, 72, 1-9.

Echevarría, J. (2007). Aceleraciones en Telépolis. *Contrastes: Revista Cultural*, 47, 131-137.

Elliot, A. & Dweck, C. (Eds.) (2005). *Handbook of Competence and Motivation*. New York: The Guilford Press.

Fluck, A., & Dowden, T. (2013). On the Cusp of Change: Examining Pre-service Teachers' Beliefs about ICT and Envisioning the Digital Classroom of the Future. *Journal of Computer Assisted Learning*, 29 (1), 43-52. doi: 10.1111/j.1365-2729.2011.00464.x

Gallego, D. J. & Alonso C. (Eds.). (2000). *La Informática en la Práctica Docente*. Madrid: Edelvives.

Goktas, Y., Yildirim, S., & Yildirim, Z. (2008). A Review of ICT Related Courses in Pre-Service Teacher Education Programs. *Asia Pacific Education Review*, 9, 168-179. doi: 10.1007/bf03026497

Goktas, Y., Yildirim, S., & Yildirim, Z. (2009). Main Barriers and Possible Enablers of ICTs Integration into Pre-Service Teacher Education Programs. *Educational Technology & Society*, 12(1), 193-204. doi: 10.1007/bf03026497

Gómez-Galán, J. & Lacerda, G. (Eds.) (2012). *Informática e Telemática na Educação.* 2 vols. Brasilia: Liber Livro.

Gómez-Galán, J. & Mateos, S. (2002). Versatile Spaces for the Use of the Information Technology in Education. In N. Mastorakis (Ed.) *Advances in Systems Engineering, Signal Processing and Communications* (pp. 351–361). New York: WSEAS Press.

Gómez-Galán, J. & Mateos, S. (2004). Design of Educational Web Pages. *European Journal of Teacher Education*, 17 (1), 99-107. doi: 10.1080/0261976042000211793

Gómez-Galán, J. (2003). *Educar en Nuevas Tecnologías y Medios de Comunicación*. Seville: Fondo Educación CRE.

Gómez-Galán, J. (2007). Los Medios de Comunicación en la Convergencia Tecnológica: Perspectiva Educativa. *Comunicación y Pedagogía: Nuevas Tecnologías y Recursos Didácticos*, 221, 44-50.

Gómez-Galán, J. (2011). New Perspectives on Integrating Social Networking and Internet Communications in the Curriculum. *eLearning Papers*, 26, 1-7.

Gómez-Galán, J. (2014). Transformación de la Educación y la Universidad en el Postmodernismo Digital: Nuevos Conceptos Formativos y Científicos. In F. Durán (ed.) *La Era de las TIC en la Nueva Docencia* (pp. 171-182). Madrid: McGraw-Hill

Grosskopf, K. K. (2009). *Exploring the Complexities of Learning Motivation in Pre-Service Teacher Education Students: A Grounded Theory Approach* (Unpublished doctoral dissertation). University of Nebraska, Lincoln.

Hammond, M. (2014). Introducing ICT in schools in England: Rationale and consequences. *British Journal of Educational Technology*, 45(2), 191-201. doi: 10.1111/bjet.12033

Haydn, T. A., & Barton, R. (2007). Common Needs and Different Agendas: How Trainee Teachers Make Progress in Their Ability to Use ICT in Subject Teaching. Some Lessons from the UK. *Computers & Education*, 49, 1018-1036. doi: 10.1016/j.compedu.2005.12.006

Hiralaal, A. (2013). ICT in Practice at the Durban University of Technology. In *Proceedings of the 8Th International Conference on E-Learning* (pp. 176-184). Cape Town: ACPIL.

Hortovanyi, L. & Ferincz, A. (2015). The Impact of ICT on Learning on-the-Job. *The Learning Organization*, 22 (1), 2-13. doi: 10.1108/TLO-06-2014-0032

Juszczyk, S. (2006). Education in the Knowledge-based Society - Chosen Aspects. *The New Educational Review*, 10, 15-31.

Kennedy, G. E., Judd, T. S., Churchward, A., Gray, K. & Krause, K. L. (2008). First Year Students' Experiences with Technology: Are they Really Digital Natives? *Australasian Journal of Educational Technology*, 24 (1), 108-122. doi: 10.14742/ajet.1233

Khan, S. H., & Hasan, M. (2013). Introducing ICT into Teacher-Training Programs: Problems in Bangladesh. *Journal of Education and Practice*, 4(14), 79-86.

Kim, H., Choi, H., Han, J., & So, H. J. (2012). Enhancing Teachers' ICT Capacity for the 21st Century Learning Environment: Three Cases of Teacher Education in Korea. *Australasian Journal of Educational Technology*, 28, 965-982. doi: 10.14742/ajet.805

Lanier, J. (2006). *Information is an Alienated Experience*. New York: Basic Books.

Leask, M., & Pachler, N. (Eds.). (2013). *Learning to Teach Using ICT in the Secondary School: A Companion to School Experience*. London: Routledge.

Li, L. & Walsh, S. (2011). Technology Uptake in Chinese EFL Classes. *Language Teaching Research*, 15 (1), 99-125. doi: 10.1177/1362168810383347

López-Meneses, E. & Gómez-Galán, J. (2010). Prácticas Universitarias Constructivistas e Investigadoras con Software Social. *Praxis*, 5, 23-45.

Martín, J., Beltrán, J., & Pérez, L. (coord.) (2003). *Cómo Aprender con Internet*. Madrid: Fundación Encuentro.

Montaser, L., Mortada, M., & Fawzy, S. (2012). The Role of Education in Development. In *Proceedings EDULEARN12* (pp. 3901-3907). Barcelona: IATED.

Mueller, J., Wood, E., Willoughby, T., Ross, C., & Specht, J. (2008). Identifying Discriminating Variables between Teachers who Fully Integrate Computers and Teachers with Limited Integration. *Computers & Education*, 51, 1523-1537. doi: 10.1016/j.compedu.2008.02.003

Murray, D. W. & Rabiner, D. L. (2014). Teacher Use of Computer-Assisted Instruction for Young Inattentive Students: Implications for Implementation and Teacher Preparation. *Journal of Education and Training Studies*, 2 (2), 58-66. doi: 10.11114/jets.v2i2.283

Pattaro, C. (2015). New Media & Youth Identity. Issues and Research Pathways. *Italian Journal of Sociology of Education*, 7 (1), 297-327

Pećanac, R., Lambić, D., & Marić, M. (2011). The Influence of the Use of Educational Software on the Effectiveness of Communication Models in Teaching. *The New Educational Review*, 26, 60-70.

Peeraer, J. & Van Petegem, P. (2012). The Limits of Programmed Professional Development on Integration of Information and Communication Technology in Education. *Australasian Journal of Educational Technology*. 28, 1039-1056. doi: 10.14742/ajet.809

Petrauskiene, R. & Volungeviciene, A. (2006). Teacher Motivation Factors to Use ICT in Teaching. In V. Dagiene & R. Mittermeir (Ed.). *Information Technologies at School* (pp.329-339). Vilnius: Publishing House.

Pino, M. & Soto, J. (2010). Identificación del Dominio de Competencias Digitales en el Alumnado del Grado de Magisterio. *Teoría de la Educación. Educación y Cultura en la Sociedad de la Información*, 11 (3), 336-362.

Rogers, L. & Twidle, J. (2013). A Pedagogical Framework for Developing Innovative Science Teachers with ICT. *Research in Science & Technological Education*, 31, 227-251. doi: 10.1080/02635143.2013.833900

Schunk D.H. & Zimmerman, B.J. (Eds.) (2008). *Motivation and Self-Regulated Learning: Theory, Research, and Applications*. New York: Lawrence Erlbaum Associates.

Selwyn, N (2006), The Use of Computer Technology in University Teaching and Learning: A Critical Perspective. *Journal of Computer Assisted Learning*, 23, 83-94. doi: 10.1111/j.1365-2729.2006.00204.x

Sime, D., & Priestley, M. (2005). Student Teachers' first Reflections on Information and Communications Technology and Classroom Learning: Implications for Initial Teacher Education. *Journal of Computer Assisted Learning*, 21, 130-142. doi: 10.1111/j.1365-2729.2005.00120.x

Soler-Costa, R. (2011). Bloom's Taxonomy in the Digital Era: Development of Teaching-Learning Processes. In *Proceedings INTED2011* (pp. 6651-6660). Valencia: INTED.

Tondeur, J., Roblin, N. P., van Braak, J., Fisser, P., & Voogt, J. (2013). Technological Pedagogical Content Knowledge in Teacher Education: In Search of a New Curriculum. *Educational Studies*, 39(2), 239-243. doi: 10.1080/03055698.2012.713548

Tondeur, J., van Braak, J., Sang, G., Voogt, J., Fisser, P., & Ottenbreit-Leftwich, A. (2012). Preparing Pre-Service Teachers to Integrate Technology in Education: A Synthesis of Qualitative Evidence.

*Computers & Education*, 59, 134-144. doi: 10.1016/j.compedu.2011.10.009

Underwood, C., Parker, L., & Stone, L. (2013). Getting it Together: Relational Habitus in the Emergence of Digital Literacies. *Learning Media and Technology*, 38, 478-494. doi: 10.1080/17439884.2013.770403

UNESCO (2002). *United Nations Educational, Scientific and Cultural Organization and the World Summit on the Information Society*. Paris: UNESCO.

Van Braak, J., Tondeur, J., & Valcke, M. (2004). Explaining Different Types of Computer Use among Primary School Teachers. *European Journal of Psychology of Education*, 19(4), 407-422. doi: 10.1007/bf03173218